\newcommand{\sub}[1]{_{\mathrm{#1}}}

\documentclass[aps,twocolumn,superscriptaddress,nobalancelastpage,showpacs,
amsmath,amssymb,prl]{revtex4}

\usepackage{graphicx}
\usepackage{epstopdf}
\begin{document}

\draft

\title{Frequency and surface dependence of the mechanical loss in fused silica}

\author{Steven~D.~Penn}
\affiliation{Department of Physics, Hobart and William Smith Colleges, Geneva, 
NY 14456, USA.}
\email{penn@hws.edu}

\author{Alexander~Ageev}
\affiliation{Department of Physics, Syracuse University, Syracuse, NY 13244, 
USA.}

\author{Dan Busby}
\affiliation{LIGO Laboratory, California Institute of Technology Pasadena, CA 
91125, USA.}

\author{Gregory~M.~Harry}
\affiliation{LIGO Laboratory, Massachusetts Institute of Technology, Cambridge, 
MA 02139, USA.}

\author{Andri~M.~Gretarsson}
\affiliation{LIGO Livingston Observatory, Livingston, LA 70754, USA.}

\author{Kenji~Numata}
\affiliation{Exploration of the Universe Division, Code 663, NASA/Goddard Space 
Flight Center, Greenbelt, MD 20771, USA}

\author{Phil Willems}
\affiliation{LIGO Laboratory, California Institute of Technology Pasadena, CA 
91125, USA.}

\date{\today}

\begin{abstract}
We have compiled measurements of the mechanical loss in fused silica
from samples spanning a wide range of geometries and resonant
frequency in order to model the known variation of the loss with
frequency and surface-to-volume ratio.  
This improved understanding of the mechanical loss has
contributed significantly to the design of advanced interferometric
gravitational wave detectors, which require ultra-low loss materials
for their test mass mirrors.
\end{abstract}

\pacs{95.55.Ym, 04.80.Nn, 62.40.+i}

\maketitle

\narrowtext

\section{Introduction}

As part of the research and development for the LIGO~\cite{LIGO} and
TAMA~\cite{TAMA} gravitational wave detectors, we have conducted
investigations into the internal friction of fused silica.
Displacement of the interferometer's mirror faces arising from thermal
motion of the fused silica test mass mirrors sets a fundamental limit
to the detector sensitivity.  The frequency distribution of this noise
is directly related to the internal friction of the mirror material.

An Advanced LIGO detector has recently been proposed~\cite{AdvLIGO} with better
sensitivity than initial LIGO. The Advanced LIGO mirror
thermal noise must be as low as possible.  Two materials have been
under consideration for the mirror substrate: fused
silica and single crystal sapphire.  To its advantage, sapphire has
the higher Young's modulus and a low bulk mechanical loss ($\phi \le 3
\times 10^{-9}$)~\cite{sapphireQ}.  However, sapphire also has high
thermoelastic noise~\cite{SapphireThermo}.

In the advanced detectors, thermal noise in the mirror coatings
makes a significant contribution to the total noise
budget in the central frequency region of 30-500 Hz. Discussion on
the mechanical loss in the mirror coatings 
can be found elsewhere~\cite{GreggCoating, PennCoating, Crooks}.

Recent measurements on the mechanical loss in fused silica have
revealed a dependence on frequency~\cite{Numata, AndriThesis}
and on surface-to-volume ratio~\cite{Ageev, Penn, GreggAndri}. This
paper combines data from several of these research groups in order
to model both of these effects. The frequency dependence of the loss
agrees well with results from Weidersich {\it et
al.}~\cite{Weidersich}. In that work, loss data spanning six decades
in frequency is modeled by an asymmetric double-well potential in
the bond angle.  Together these results provide a more complete
picture of the loss in ultrapure glasses and a more physically
motivated prediction for the thermal noise in advanced
interferometric detectors. It was previously predicted 
that fused silica's loss dependence  would make it suitable for low 
frequency detectors (10 -- 100 Hz)~\cite{Riccardo}.  Indeed, this 
model's prediction of a very low mechanical loss in the LIGO 
frequency regime has motivated the recent selection of fused silica as the 
Advanced LIGO test mass substrate~\cite{DownSelect}.

\section{Theory of Loss in Fused Silica}

The thermal noise motion of the mirror surface is related to the
internal friction of the substrate by the fluctuation-dissipation
theorem~\cite{Callen52}.  The internal friction of very pure fused
silica is associated with strained Si-O-Si bonds, where the energy of
the bond has minima at two different bond angles, forming an
asymmetric double-well potential.  Redistribution of the bond angles
in response to an applied strain leads to mechanical dissipation,
which at audio frequencies has a peak in the cryogenic range 20-60K.
Because fused silica is an amorphous material, there is a distribution
of potentials which must be inferred from measurements of the
dissipation.  It can be shown~\cite{Weidersich} that the frequency
dependence of the loss should exhibit a power law spectrum with
exponent $k_BT/V_0$ at low temperatures.  Both this power law, with
$V_0/k_B=319K$, and the distribution of potentials have been
measured~\cite{Weidersich}. The power law exponent of a
relaxation process cannot exceed 1, and is expected to saturate
near 300 K.  At room temperature the exponent is
0.76. 

At elevated temperatures there is another loss peak arising from a
double-well potential associated with the Si-O-Si bond angles.  For
this peak the bond angle shift and potential barrier are much larger;
the double-well of the cryogenic loss peak is a small feature at the
minima of this larger potential well.  At room temperature, thermal
fluctuations allow the bonds to span the cryogenic double-well but not
to cross the larger potential barrier, where $V_0/k_B=3.54\times
10^4K$\cite{Bartenev}.  The calculated internal friction for this loss
peak at audio frequencies and room temperature is utterly
negligible compared to other loss mechanisms cited herein.

\renewcommand{\arraystretch}{1.4}

A separate loss mechanism
exists in the surface of the glass.  The contribution from the surface 
loss depends on the mode of the sample. The total energy lost per oscillation 
in an isotropic sample undergoing slowly decaying vibration, can be described 
by the integral of the local loss angle,$\phi(\vec{r})$ with the energy density
$\rho_{\scriptscriptstyle E}(\vec{r})$
\begin{equation}\label{EnergyLostPerCycle}
\begin{array}{l}
\Delta E = 2 \pi \int_{\mathcal V} \rho_{\scriptscriptstyle
E}(\vec{r})\,\phi(\vec{r})\,d^3r
\end{array}
\end{equation}
where ${\mathcal V}$ is the sample volume. Assuming that the local loss 
angle is constant and equal to $\phi\sub{bulk}$ everywhere except within 
a distance $h$ of the surface, and that the energy density in that 
surface layer is approximately the energy density at the surface, then 
the loss can be expressed as~\cite{AndriThesis}  
\begin{equation}\label{GeneralPhiExpression}
\begin{array}{l}
\phi=\phi\sub{bulk} +  \mu\,\alpha_{\mathrm{s}}\frac{S}{V}
\end{array}
\end{equation}
where $S$ is the surface area of the sample and $\mu$ is a factor of
order unity that depends on the mode shape. 
The surface loss parameter, $\alpha_{\mathrm{s}}$, is typically 
several picometers for flame polished or flame drawn fused silica 
but much higher for abrasively polished surfaces.

\renewcommand{\arraystretch}{1}

\section{Experimental Method}

The measurements at Syracuse University
(SU)~\cite{GreggAndri,AndriThesis,Penn,Ageev} were performed on
fiber/rod samples with diameters ranging from 0.1 -- 8 mm over
resonant frequencies less than 5 kHz.  The samples were drawn from and 
left attached to a massive bob of Suprasil~\cite{heraeus}, thus 
forming a cantilever beam. This bob was welded to a vibration 
isolating suspension formed by similar silica bobs connected by
thin silica fibers. In  a vacuum of
$\approx 10^{-6}$ torr, the samples were made resonant by an
electrostatic comb exciter, and their position was measured using a
shadow sensor.

The measurements at University of Tokyo~\cite{Numata} were performed
on cylindrical samples with optically polished surfaces. The diameters
and  heights were 70 mm and 60 mm, respectively. The samples were
annealed in a vacuum furnace. To exclude the support loss, the samples
were supported at nodal points of their vibrational modes during the $Q$
measurements.

The Caltech measurements were performed on a spare input test mass
for the initial LIGO interferometers, a superpolished right cylinder made from
Suprasil~312 with a diameter of 25.4~cm and a thickness of 10~cm. It
was suspended in a $\approx 10^{-6}$~Torr vacuum by a loop of
polished stainless steel wire greased with lard. The elastic modes
of the mass were excited with an electrostatic actuator and the mode
amplitude was monitored using a birefringence sensor. Since friction
at the wire could reduce $Q$, only modes with small motion at the
point of wire contact were used in the fit.

\section{\label{sec:modeling}Modeling Method}

Resonant $Q$ measurements from each of the labs were submitted for
generating this model of the loss.  The measurements spanned several
types of fused silica, V/S ratios from 0.03 -- 28 mm, and frequency
up to $10^{5}$ Hz.  The data was first separated by silica type
since the loss is known to vary significantly between varieties of
fused silica~\cite{Startin, Ageev, Numata}.  Only Suprasil 2 and 312
had sufficient data to warrant a fit over both frequency and V/S
ratio. Characteristics of these samples are listed in
Table~\ref{tbl:data}.

\begingroup
\squeezetable
\begin{table}[htb]
\begin{ruledtabular}
	 \begin{tabular}{|c|c|c|c|c|c|c|c|c|}
		  \hline
		  \textbf{Label} & \textbf{Type} & $\mathbf{D}$ &
		  $\mathbf{h}$ & $\mathbf{V/S}$ & \textbf{Surface} & \textbf{Anneal} &
		  \textbf{Lab}  \\
		  \hline \hline
		  P1 & 312, cyl. & 254 & 100 & 28 & SP & None & Caltech  \\
		  \hline
		  K12 & 312, cyl. & 70 & 60 & 11 & SP & 980$^{o}$C, vac. & Tokyo  \\
		  \hline
		  SU2 & 312, rod & 3 & -- & 0.75 & FP & 1025$^{o}$C in Ar & SU  \\
		  \hline
		  SV4 & 312, rod & 8 & -- & 2 & FP & 950$^{o}$C in Ar & SU  \\
		  \hline \hline
		  AG5 & 2, rod & 3.5 & 188 & 0.88 & FP & None & SU  \\
		  \hline
		  AH1 & 2, rod & 0.300 & 108 & 0.075 & FP & None & SU  \\
		  \hline
		  AN1 & 2, rod & 0.318 & 160 & 0.080 & FP & None & SU  \\
		  \hline
		  AB1 & 2, rod & 0.062 & 175 & 0.016 & FP & None & SU  \\
		  \hline
		  AC1 & 2, rod & 0.340 & 310 & 0.085 & FP & None & SU  \\
		  \hline
		  AF1 & 2, rod & 0.120 & 130 & 0.030 & FP & None & SU  \\
		  \hline
		  K13 & 2, cyl. & 70 & 60 & 11 & SP & 900$^{o}$C, vac. & Tokyo  \\
		  \hline
	 \end{tabular}
	 \centering
	 \caption{Sample characteristics: \textit{Type} lists Heraeus Suprasil 
    variety and shape. Samples are cylinders with diameter (D), height (h), 
    and volume-to-surface ratio (V/S) given in mm. Surface types are 
    superpolished (SP) and flame polished (FP).  }
	 \label{tbl:data}
	 \end{ruledtabular}
\end{table}
\endgroup

We chose a model for the mechanical loss that included terms
describing the frequency dependence, the surface loss, and the
thermoelastic loss.  The loss function took the form:

\begin{eqnarray}
	 \phi(f,\frac{V}{S}) & = & \phi_{\mathrm{surf}} + \phi_{\mathrm{bulk}}
	 + \phi_{\mathrm{th}}\\
	  & = & C_{1}(\frac{V}{S})^{-1} + C_{2} (f/1~\mathrm{Hz})^{C_{3}} +
	C_{4} \phi_{\mathrm{th}}
	 \label{eq:model}
\end{eqnarray}

\noindent where $C_{1}=\mu\alpha_{\mathrm{s}}$ from
Eqn.~\ref{GeneralPhiExpression}. Given that the surface loss term only
contributes significantly to the rod (fiber) samples, we have assumed
for all samples that $\mu \approx 2$ which is appropriate for
cylindrical rods. We have also not distinguished the loss angle arising
from the Young's modulus from that due to the shear modulus.

\begin{figure}[htbp]
	 \centering
	\includegraphics[width=8.75cm]{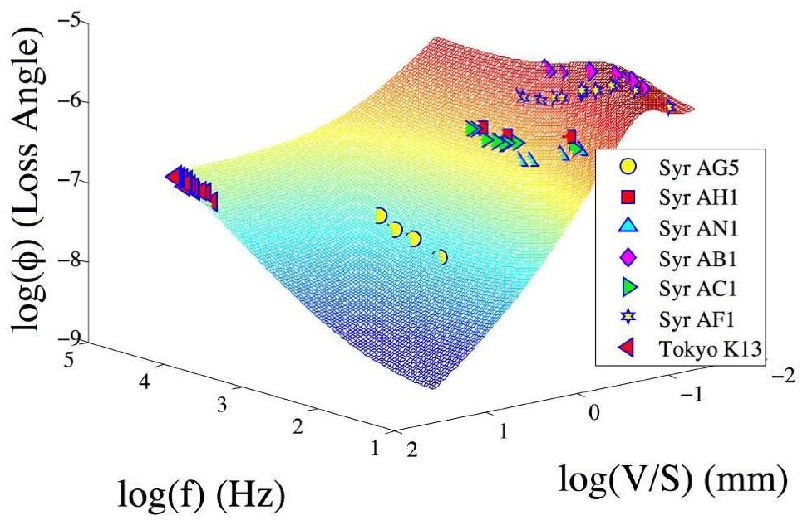}
	\includegraphics[width=8.75cm, height=6cm]{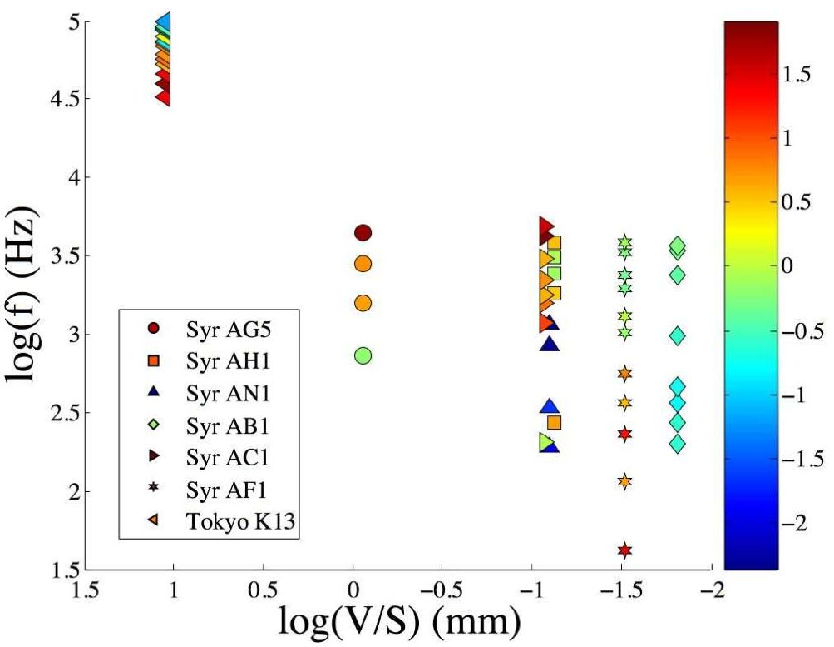}
	 \caption{Suprasil 2 mechanical loss data: Best fit surface 
	 (upper) and Deviation in units of sample variance (lower). }
	 \label{fig:S2fit}
\end{figure}

\begin{figure}[htbp]
	\centering
	\includegraphics[width=8.75cm]{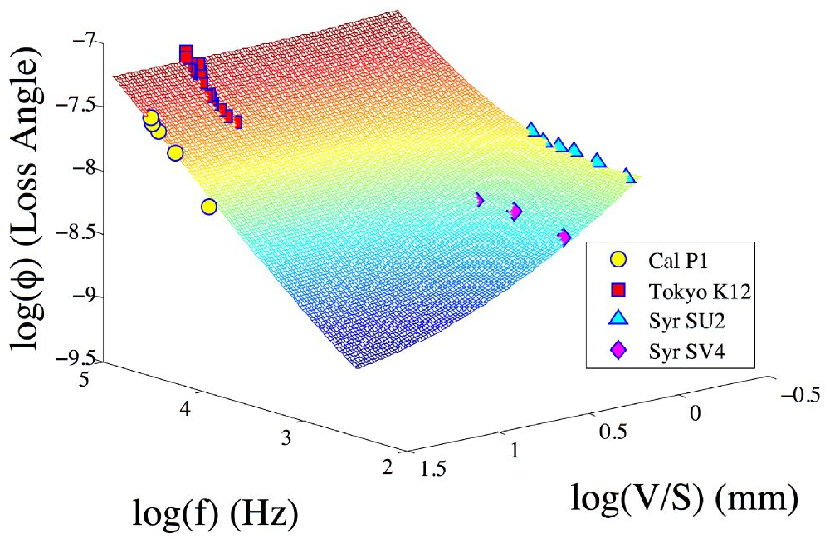}
	\includegraphics[width=8.75cm,height=6cm]{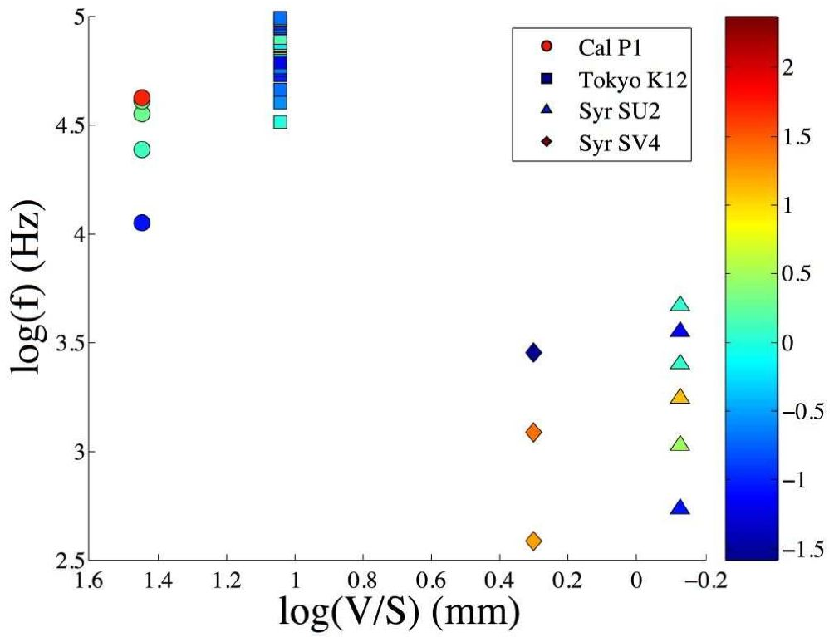}
	\caption{Suprasil 312 mechanical loss data: Best fit surface 
	(upper) and Deviation in units of sample variance (lower). }
	\label{fig:S312fit}
\end{figure}

\begingroup
\squeezetable
\begin{table}[htb]
\begin{ruledtabular}
    \begin{tabular}{|c|c|c|c|c|}
        \textbf{Type} & $\mathbf{C_{1}}$ (pm) & 
        $\mathbf{C_{2}}$ ($\times 10^{-11}$) &
        $\mathbf{C_{3}}$ & $\mathbf{C_{4}}$  \\
        \hline \hline
        
        2 & $12.1 \pm 0.8$ & $1.18 \pm 0.04$ & $0.77 \pm 0.02$ & $0.61 \pm 0.05$  \\
        \hline
        312 & $6.5 \pm 0.2$ & $0.76 \pm 0.02$ & $0.77 \pm 0.02$ &   \\

    \end{tabular}
    \centering
    \caption{Fit coefficients for Suprasil 2 and 312. }
    \label{tbl:fitcoefs}
    \end{ruledtabular}
\end{table}
\endgroup

The thermoelastic loss, $\phi_{\mathrm{th}}$, which is negligible in all but the
thinnest fiber samples, is described for fibers by:

\begin{equation}\label{eq:thermoelastic}
	 \begin{array}{l}
		  \phi_{\mathrm{th}} = \frac{Y \alpha^{2}T}{\rho\, C_{\mathrm{m}}}
		  \frac{2 \pi f \tau}{1+(2 \pi f \tau)^{2}}\\
		  \\
		  \tau = (d^{2}\,\rho\,C_{\mathrm{m}})/(13.55\,\kappa)
	 \end{array}
\end{equation}

\noindent where $Y$ is the Young's modulus, $\alpha$ is the coefficient of
thermal expansion, $T$ is temperature, $\rho$ is the density, $C_{\mathrm{m}}$ 
is the mass
specific heat capacity, $d$ is the diameter, and $\kappa$ is the
thermal conductivity. We fit the amplitude of $\phi_{\mathrm{th}}$ to
account for small changes in the coefficient of thermal expansion
among samples.  Variations in the fiber diameter can also slightly
alter the shape of the thermoelastic peak.  Neither of these effects
significantly affect the frequency or surface loss terms.

Measurements of large resonant $Q$'s are subject to numerous 
mechanisms that can greatly reduce the $Q$ and few processes that can
 increase it.  These  effects produce a distribution
in the systematic error that is asymmetric, heavily skewed toward
lower $Q$, and unique for each experiment.  Standard data analysis
techniques based on normally distributed error, such as linear least
squares (LLS) fitting, are therefore inappropriate for analyzing our
full data set.  We circumvent this problem by first limiting our data
to the best measurement at each $(f, V/S)$ point for each sample.  A LLS 
fitting routine is applied with the sample variance approximating the 
actual variance of the data.  This method is
commonly used in analyzing mechanical loss measurements where the lowest
loss measurement closely approximates  actual mechanical
loss for a sufficiently large set of measurements~\cite{GreggAndri, Penn}.
The results of the method are displayed in
Figure~\ref{fig:S2fit} for Suprasil 2 and in Figure~\ref{fig:S312fit}
for Suprasil 312. The fit coefficients are listed in
Table~\ref{tbl:fitcoefs}. The frequency dependence, $C_{3}$, 
agrees well with results from Weidersich {\it et
al.}~\cite{Weidersich}. The thermoelastic amplitude, $C_{4}$, is 
similar to earlier measurements~\cite{AndriThesis}. Assuming no unforeseen
loss mechanisms, the Advanced LIGO test masses ($V/S \approx 40$ mm) 
have a predicted loss ($\phi(\mathrm{100\ Hz}) \approx 4 \times 10^{-10}$) 
that is a several-fold improvement over previous estimates.  

\begin{figure}[h]
\centering
\includegraphics[width=8.75cm]{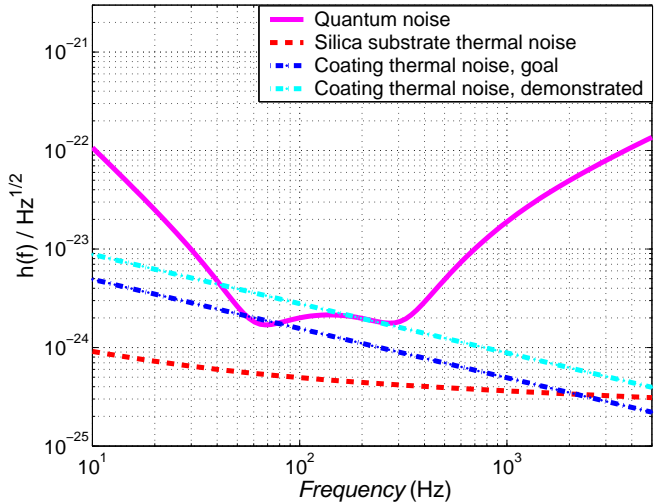}
\caption{Estimated Advanced LIGO thermal noise for a Suprasil 312 test mass 
substrate and for two mirror coatings: the best measured and the 
research goal. Laser quantum noise provided for comparison.}
\label{fig:bench1}
\end{figure}

\begingroup
\squeezetable
\begin{table}[htb]
\begin{ruledtabular}
\begin{tabular}{|c|c|c|c|}
Coating Loss & Loss Angle $\phi_{||}$ & BNSI Range  & BH/BH Range  \\
\hline
Measured & $1.6 \times 10^{-4}$ & 190 MPc & 840 MPc\\
Goal & $5.0 \times 10^{-5}$ & 230 MPc& 1060 MPc\\
\end{tabular}
\centering \caption{The distance a single Advanced LIGO interferometer could 
detect a neutron star or 10 $M_{\odot}$ black hole binary inspiral, 
assuming a Suprasil 312 test mass and two possible mirror coatings: the best measured 
and the research goal.}
\label{tbl:reach}
\end{ruledtabular}
\end{table}
\endgroup

\section{Implications for Advanced Detectors}

The low mechanical loss of silica in the 10 -- 1000~Hz bandwidth,
coupled with its optical and thermal properties, makes it an
attractive material for the optics of next generation interferometric
gravitational wave detectors.  Fused silica has recently been chosen as the
test mass substrate for Advanced LIGO~\cite{AdvLIGO}, which has been
approved and recommended for funding by the US National Science
Foundation.

If the bulk and surface loss predicted herein can be achieved, the
mirror thermal noise in Advanced LIGO with fused silica mirrors will likely
be dominated by the coating~\cite{Levin, GreggCoating,PennCoating}.
The mirror thermal noise contributions to the total Advanced LIGO
noise budget are shown in Figure~\ref{fig:bench1}.
Table~\ref{tbl:reach} shows the predicted sensitivity of Advanced LIGO
with silica optics to two possible sources of gravitational waves:
binary neutron star inspirals (BNSI) and binary
10 $M_{\odot}$ black hole inspirals.  Two
different scenarios of coating thermal noise are shown: the best 
measurements to date~\cite{GreggOttawa} and the research goal.  The
sensitivity goal for a single Advanced LIGO interferometer is to
observe BNSI, averaged over sky position and
polarization, to a distance of $\approx$ 200~Mpc.  (See
Harry~\cite{HarrySphere} for a description of a LIGO range
calculation.)  

\section{Conclusions}
We have shown that the
mechanical loss of fused silica can be described by a model
that includes surface loss and a frequency dependent bulk loss. The
frequency dependent loss, thought to arise from an
asymmetric double-well potential of the bond angle, agrees well with
earlier measurements~\cite{Weidersich} that spanned six decades in
frequency. This improved understanding of the loss indicates that at
large geometries and low frequency, fused silica is an excellent
material for test masses in advanced interferometric gravitational
wave detectors.

\section{Acknowlegments}
   The authors would like to thank the LIGO laboratory and LIGO
   Science Collaboration for their support and review of this
   work.  This research was supported by the National Science
   Foundation under cooperative agreements PHY-9210038 \& PHY-0107417 (LIGO
   laboratory) and awards PHY-9801158 \& PHY-0098715 (Caltech), 
   PHY-0245118 \& PHY-0355118 (HWS), and PHY-0140335 (SU).

\end{document}